\documentclass{PoS}

\PoS{PoS(LAT2005)026}

\title{The Exotic Baryon $\Theta^+(1540)$ on the Lattice }

\ShortTitle{The Exotic Baryon $\Theta^+(1540)$ on the Lattice}

\author{\speaker{Dieter Hierl}\\
        University of Regensburg\\
        E-mail: \email{dieter.hierl@physik.uni-regensburg.de}}

\author{Christian Hagen\\
        University of Regensburg\\
        E-mail: \email{christian.hagen@physik.uni-regensburg.de}}

\author{Andreas Sch\"afer\\
        University of Regensburg\\
        E-mail: \email{andreas.schaefer@physik.uni-regensburg.de}}

\abstract{We report on a study of the pentaquark $\Theta^+(1540)$, using a variety of different interpolating fields. We use
Chirally Improved fermions in combination with Jacobi smeared quark sources to improve the signal and get reliable results
even for small quark masses. The results of our quenched calculations, which have been done on a $12^3\times24$ lattice with
a lattice spacing of $a=0.148~{\rm fm}$, do not provide any evidence for the existence of a $\Theta^+$ with positive parity.
We do observe, however, a signal compatible with nucleon-kaon scattering state. For the negative parity the results are
inconclusive, due to the potential mixture with nucleon-kaon and $N^*$-kaon scattering states.}

\FullConference{XXIIIrd International Symposium on Lattice Field Theory\\
		 25-30 July 2005\\
		 Trinity College, Dublin, Ireland}
		 
\newlength{\myVSpace}
\newcommand\xstrut{\raisebox{.5\myVSpace}{\rule{0pt}{\myVSpace}}}

\begin{document}
\section{Introduction}
\label{intro}\noindent
The possible discovery of the $\Theta^+(1540)$ by the LEPS Collaboration at SPring-8 \cite{Nakano:2003qx} has initiated
great interest in exotic baryons. Since then, there has been a large number of experiments that have confirmed this result, e.g.
\cite{Stepanyan:2003qr,Barth:2003es}, but also about the same number that could not confirm it, e.g.
\cite{Bai:2004gk,Knopfle:2004tu}.\newline
To confirm or disprove the existence of the $\Theta^+$, if a clear conclusion of the lattice community could be reached and
would eventually be experimentally confirmed, this gave a substantial boost to lattice QCD. Therefore, many groups have
started to work on this problem. Here we present our first results.\newline
In our calculations we perform a qualitative study using different types of spin-$\frac{1}{2}$ operators with the quantum
numbers of the $\Theta^+$. We compute all cross correlators and use the variational method
\cite{Michael:1985ne,Luscher:1990ck} to extract the lowest lying eigenvalues. These are used to create effective mass plots
for a comparison to the $N$-$K$ scattering state which we computed separately on the the same lattice.
\section{Details of the calculation}
\label{settings}\noindent
We considered the following interpolating fields as basis for our correlation matrix:
\begin{itemize}
\item
Currents suggested by Sasaki \cite{Sasaki:2003gi}:
\begin{eqnarray}
\label{sasaki1}
\Theta_+^1&=&\epsilon_{abc}\epsilon_{aef}\epsilon_{bgh}
(u_e^T C d_f)(u_g^T C \gamma_5 d_h )C\bar{s}_c^T
\\\nonumber\\
\label{sasaki2}
\Theta_{+,\mu}^2&=&\epsilon_{abc}\epsilon_{aef}\epsilon_{bgh}
(u_e^T C \gamma_5 d_f)(u_g^T C \gamma_5 \gamma_\mu d_h )C\bar{s}_c^T
\\\nonumber\\
\label{sasaki3}
\Theta_{-,\mu}^3&=&\epsilon_{abc}\epsilon_{aef}\epsilon_{bgh}
(u_e^T C d_f)(u_g^T C \gamma_5 \gamma_\mu d_h )C\bar{s}_c^T
\end{eqnarray}
\item
A current which is a suggestion by L. Ya. Glozman \cite{Glozman:2003sy}, however, using only s-wave quarks instead of a
mixture of s-wave and p-wave quarks in (\ref{glozman2}):
\begin{eqnarray}
\label{glozman2}I_{\mu}&=&(\delta_{ae}\delta_{bg}+\delta_{be}\delta_{ag})\epsilon_{gcd}\\
&\times&\left(
\begin{array}{c}
u_a^T C u_b\\
\frac{1}{\sqrt{2}} (u_a^T C d_b + d_a^T C u_b )\\
d_a^T C d_b
\end{array}
\right)\nonumber\\
&\times&\left(
\begin{array}{c}
d_c^T (C \gamma_\mu ) d_d\\
\frac{1}{\sqrt{2}} (u_c^T (C \gamma_\mu ) d_d + d_c^T (C \gamma_\mu ) u_d)\\
u_c^T (C \gamma_\mu ) u_d
\end{array}
\right)
C\bar{s}_e^T\nonumber
\end{eqnarray}
\item
An other current related to suggestions by L. Ya. Glozman \cite{Glozman:2003sy}, using only s-wave quarks and a factor
$(\delta_{ae}\delta_{bg}-\delta_{be}\delta_{ag})$ instead the factor $(\delta_{ae}\delta_{bg}+\delta_{be}\delta_{ag})$ in
(\ref{glozman4}):
\begin{eqnarray}
\label{glozman4}I_{\nu}&=&(\delta_{ae}\delta_{bg}-\delta_{be}\delta_{ag})\epsilon_{gcd}\times\\
&\times&\left(
\begin{array}{c}
u_a^T (C \sigma_{\mu \nu} ) u_b\\
\frac{1}{\sqrt{2}} (u_a^T (C \sigma_{\mu \nu} ) d_b + d_a^T (C \sigma_{\mu \nu} ) u_b )\\
d_a^T (C \sigma_{\mu \nu} ) d_b
\end{array}
\right)\nonumber\\
&\times&\left(
\begin{array}{c}
d_c^T (C \gamma_\mu ) d_d\\
\frac{1}{\sqrt{2}} (u_c^T (C \gamma_\mu ) d_d + d_c^T (C \gamma_\mu ) u_d)\\
u_c^T (C \gamma_\mu ) u_d
\end{array}
\right)
C\bar{s}_e^T\nonumber
\end{eqnarray}
\end{itemize}
In order to get only spin-$\frac{1}{2}$ pentaquarks, we have to project the spin. This is done using the spin projection
operator for a Rarita-Schwinger field \cite{Leinweber:2004it}. We also apply a parity projection to be able
to distinguish both parity channels. The parity channel of the $\Theta^+$ is not known. There are conflicting theoretical
predictions and no conclusive experimental data.\newline
For the interpolators (\ref{glozman2}) and (\ref{glozman4}) p-wave quarks are required. Since we do not have p-wave sources
we had to adjust the color structure in the interpolator (\ref{glozman4}) to obtain a signal at all.\newline
The interpolators (\ref{glozman2}) and (\ref{glozman4}) are linear combinations of two diquarks with $I=1$. Thus they are a
mixture of isospin $I=0$ and $I=2$ states. The interpolators (\ref{sasaki1}), (\ref{sasaki2}) and (\ref{sasaki3}) do not
have any visible isospin projection, but hidden ones.\newline
Then we use the five interpolators to calculate a cross correlation matrix $C_{ij}(t)$ which is then inserted into
the generalized eigenvalue problem
\begin{eqnarray}
C_{ij}(t) \vec{v}_i^{(k)}&=&\lambda^{(k)}(t) C_{ij}(t_0) \vec{v}_i^{(k)}.
\end{eqnarray}
The solutions of this equation behave like
\begin{eqnarray}
\lambda^{(k)}(t)&\propto&\exp\left(-m^{(k)}(t-t_0)\right).
\end{eqnarray}
These eigenvalues are used to compute effective masses according to
\begin{eqnarray}
m_{eff}(t)=\ln\left(\frac{\lambda(t)}{\lambda(t+1)}\right).
\end{eqnarray}
Ordering the five eigenvalues according to their absolute value the largest eigenvalue in the positive parity channel should
give the $\Theta^+$ mass if $\Theta^+$ is a positive parity particle. The second largest eigenvalue in the negative parity
channel should give the $\Theta^+$ mass, where the largest eigenvalue corresponds to the a $N$-$K$ scattering state at
rest.\newline
In our quenched calculation we use the Chirally Improved Dirac operator \cite{Gattringer:2000js,Gattringer:2000qu}. It is an
approximate solution of the Ginsparg-Wilson equation \cite{Ginsparg:1981bj}, with good chiral behavior
\cite{Gattringer:2003qx,Gattringer:2002sb}. The gauge fields are generated with the L\"uscher-Weisz gauge
action \cite{Luscher:1984xn,Curci:1983an} at $\beta = 7.90$. The corresponding value of the lattice spacing is $a=0.148~{\rm
fm}$ as determined from the Sommer parameter in \cite{Gattringer:2001jf}. The strange quark mass is fitted using the
pseudoscalar $K$ meson. The error bars are computed using the jackknife method. The parameters of our calculation are collected in Table
\ref{latticetable}.
\begin{table}
\begin{center}
\begin{tabular}{|c|c|}
\hline
size $L^3\times T$&$12^3\times24$\xstrut\\\hline
$a~[{\rm fm}]$&0.148\\\hline
$L~[{\rm fm}]$&$\approx1.8$\\\hline
$\#$conf N&100\\\hline
&0.02, 0.03, 0.04, 0.05,\\
quark masses $am_q$&0.06, 0.08, 0.10, 0.12,\\
&0.16, 0.20\\\hline
smearing parameters:&$n=18$, $\kappa=0.210$\\\hline
s-quark mass $am_s$&0.0888(17)\\\hline
\end{tabular}
\end{center}
\caption{Parameters of our calculations.}
\label{latticetable}
\end{table}
\section{Results}
\label{results}
\begin{figure*}[t]
\begin{center}
\resizebox{0.6\textwidth}{!}{\includegraphics{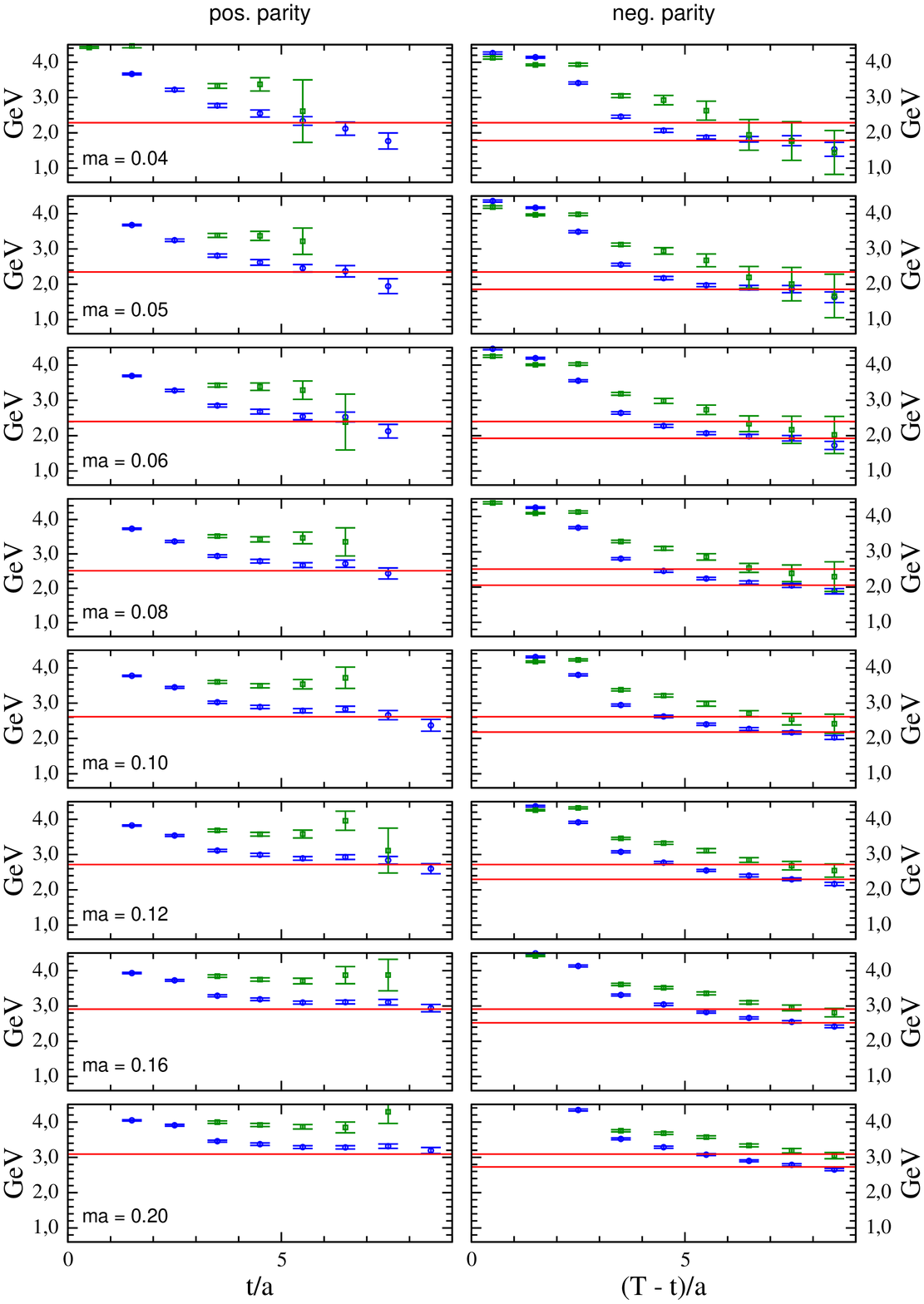}}
\end{center}
\caption{Results from cross-correlation of five interpolators. We use Jacobi smeared Gaussian quark sources for all our
quarks. Only the effective masses of the two largest eigenvalues are plotted.}
\label{fig:1}
\end{figure*}
\noindent The results of our calculations are shown in Fig. \ref{fig:1}, where we plot the effective masses of the two
lowest lying states of both parity channels obtained with the cross-correlation technique. These states are approaching a
possible plateau very slowly as we expected, since states consisting of five quarks are very complicated and therefore
should contain a large number of excited states which have to die out before the effective mass reaches a plateau. We use in
addition to the cross-correlation technique Jacobi smeared Gaussian quark sources for all our quarks to improve the signal
for the lowest lying states.\newline
The lower horizontal line in the negative parity channel is the sum of the nucleon and kaon mass at rest in the ground state
obtained from a separate calculation on the same lattice. Since we project the final state to zero momentum a scattering
state can also be a two particle state where the two particles have the same but antiparallel momentum, i.e.
$\vec{p_N}=-\vec{p_K}$. We use the relativistic $E$-$p$-relation to calculate the energy of such
states,
\begin{eqnarray}
\label{relep}
E&=&\sqrt{p^2+m_N^2}+\sqrt{p^2+m_K^2},
\end{eqnarray}
where the smallest momentum is $2\pi/L\approx700~{\rm MeV}$ on our lattice. In Fig. \ref{fig:1} this energy is represented by the upper
horizontal line.\newline
We find effective mass plateaus which are consistent with $N$-$K$ scattering states in the negative parity channel as we
expected. We find that the second state is noisy but within errors consistent with the energy in (\ref{relep}). Therefore, it
is most likely that we do not observe a $\Theta^+$ state in the negative parity channel. However, this conclusion is not
completely certain, if the $\Theta^+$ is broad, because such a $\Theta^+$ state would mix strongly with the continuum
states.\newline
In the positive parity channel one expects to find either a bound $\Theta^+$ or an excited $N$-$K$ scattering state. For
such an excited state there are several possibilities, e.g., $N^*$-$K$, or $N$-$K$ with a relative angular momentum, and so
on.\newline
On the positive parity side, we also show the two lowest lying states obtained from our calculation. Both of them
are too heavy to describe a $\Theta^+$ state. They probably correspond to excited $N$-$K$ scattering states.
If there were a signal belonging to the $\Theta^+$ it is supposed to lie below the red line assuming that the chiral
extrapolation of the $\Theta^+$ does not lead to dramatic effects below our smallest quark mass.
\section{Conclusion}
\label{conclusion}\noindent
In this article we present the results of a pilot study of the $\Theta^+$ using different types of operators. We find that
for negative parity our results are in good agreement with a $N$-$K$ scattering state in the ground state and a quite noisy
signal for the first excited state. For positive parity we find states which are typically more than $500~{\rm MeV}$ heavier than the
$\Theta^+$ and thus not compatible with a $\Theta^+$ mass of $1540~{\rm MeV}$.\footnote{We assume that the $\Theta^+$ extrapolates
smoothly in both, the chiral and the continuum limit.}\newline
Thus our calculation do not show any hints for a $\Theta^+$ in the quenched approximation with chiral fermions for
the positive channel. In the negative parity channel we would need smaller errors to be able to make a really firm statement
for the existence of a $\Theta^+$ state, but it is most likely that there is no such state in our data.
\section*{Acknowledgements}
\label{ack}\noindent
This work was funded by DFG and BMBF. We thank Ch. Gattringer, T. Burch and L.~Ya. Glozman for very helpful discussions and
their support. All computations were done on the Hitachi SR8000 at the Leibniz-Rechenzentrum in Munich, on the JUMP cluster
at NIC in J\"ulich and at the Rechenzentrum in Regensburg.
\bibliographystyle{JHEP}
\bibliography{lit}

\providecommand{\href}[2]{#2}\begingroup\raggedright\begin{thebibliography}{10}

\bibitem{Nakano:2003qx}
{\bf LEPS} Collaboration, T.~Nakano {\em et~al.}, {\it {Observation of $S=+1$
  baryon resonance in photo-production from neutron}},  {\em Phys. Rev. Lett.}
  {\bf 91} (2003) 012002, [\href{http://xxx.lanl.gov/abs/hep-ex/0301020}{{\tt
  hep-ex/0301020}}].

\bibitem{Stepanyan:2003qr}
{\bf CLAS} Collaboration, S.~Stepanyan {\em et~al.}, {\it {Observation of an
  exotic $S=+1$ baryon in exclusive photoproduction from the deuteron}},  {\em
  Phys. Rev. Lett.} {\bf 91} (2003) 252001,
  [\href{http://xxx.lanl.gov/abs/hep-ex/0307018}{{\tt hep-ex/0307018}}].

\bibitem{Barth:2003es}
{\bf SAPHIR} Collaboration, J.~Barth {\em et~al.}, {\it {Observation of the
  positive-strangeness pentaquark $\Theta^+$ in photoproduction with the SAPHIR
  detector at ELSA}},  \href{http://xxx.lanl.gov/abs/hep-ex/0307083}{{\tt
  hep-ex/0307083}}.

\bibitem{Bai:2004gk}
{\bf BES} Collaboration, J.~Z. Bai {\em et~al.}, {\it Search for the pentaquark
  state in {$\psi(2S)$} and {$J/\psi$} decays to {$K_0(S)pK^-\bar{n}$} and
  {$K_0(S)\bar{p}K^+n$}},  {\em Phys. Rev.} {\bf D70} (2004) 012004,
  [\href{http://xxx.lanl.gov/abs/hep-ex/0402012}{{\tt hep-ex/0402012}}].

\bibitem{Knopfle:2004tu}
{\bf HERA-B} Collaboration, K.~T. Knopfle, M.~Zavertyaev, and T.~Zivko, {\it
  Search for {$\Theta^+$} and{ $\Xi(3/2)^{--}$} pentaquarks in {HERA-B}},  {\em
  J. Phys.} {\bf G30} (2004) S1363--S1366,
  [\href{http://xxx.lanl.gov/abs/hep-ex/0403020}{{\tt hep-ex/0403020}}].

\bibitem{Michael:1985ne}
C.~Michael, {\it Adjoint sources in lattice gauge theory},  {\em Nucl. Phys.}
  {\bf B259} (1985) 58.

\bibitem{Luscher:1990ck}
M.~Luscher and U.~Wolff, {\it How to calculate the elastic scattering matrix in
  two- dimensional quantum field theories by numerical simulation},  {\em Nucl.
  Phys.} {\bf B339} (1990) 222--252.

\bibitem{Sasaki:2003gi}
S.~Sasaki, {\it Lattice study of exotic {$S=+1$} baryon},  {\em Phys. Rev.
  Lett.} {\bf 93} (2004) 152001,
  [\href{http://xxx.lanl.gov/abs/hep-lat/0310014}{{\tt hep-lat/0310014}}].

\bibitem{Glozman:2003sy}
L.~Y. Glozman, {\it {$\Theta^+$ in a chiral constituent quark model and its
  interpolating fields}},  {\em Phys. Lett.} {\bf B575} (2003) 18--24,
  [\href{http://xxx.lanl.gov/abs/hep-ph/0308232}{{\tt hep-ph/0308232}}].

\bibitem{Leinweber:2004it}
D.~B. Leinweber, W.~Melnitchouk, D.~G. Richards, A.~G. Williams, and J.~M.
  Zanotti, {\it Baryon spectroscopy in lattice {QCD}},
  \href{http://xxx.lanl.gov/abs/nucl-th/0406032}{{\tt nucl-th/0406032}}.

\bibitem{Gattringer:2000js}
C.~Gattringer, {\it A new approach to {Ginsparg-Wilson} fermions},  {\em Phys.
  Rev.} {\bf D63} (2001) 114501,
  [\href{http://xxx.lanl.gov/abs/hep-lat/0003005}{{\tt hep-lat/0003005}}].

\bibitem{Gattringer:2000qu}
C.~Gattringer, I.~Hip, and C.~B. Lang, {\it Approximate {Ginsparg-Wilson}
  fermions: A first test},  {\em Nucl. Phys.} {\bf B597} (2001) 451--474,
  [\href{http://xxx.lanl.gov/abs/hep-lat/0007042}{{\tt hep-lat/0007042}}].

\bibitem{Ginsparg:1981bj}
P.~H. Ginsparg and K.~G. Wilson, {\it A remnant of chiral symmetry on the
  lattice},  {\em Phys. Rev.} {\bf D25} (1982) 2649.

\bibitem{Gattringer:2003qx}
{\bf BGR} Collaboration, C.~Gattringer {\em et~al.}, {\it Quenched spectroscopy
  with fixed-point and chirally improved fermions},  {\em Nucl. Phys.} {\bf
  B677} (2004) 3--51, [\href{http://xxx.lanl.gov/abs/hep-lat/0307013}{{\tt
  hep-lat/0307013}}].

\bibitem{Gattringer:2002sb}
{\bf Bern-Graz-Regensburg} Collaboration, C.~Gattringer {\em et~al.}, {\it
  Quenched {QCD} with fixed-point and chirally improved fermions},  {\em Nucl.
  Phys. Proc. Suppl.} {\bf 119} (2003) 796--812,
  [\href{http://xxx.lanl.gov/abs/hep-lat/0209099}{{\tt hep-lat/0209099}}].

\bibitem{Luscher:1984xn}
M.~Luscher and P.~Weisz, {\it On-shell improved lattice gauge theories},  {\em
  Commun. Math. Phys.} {\bf 97} (1985) 59.

\bibitem{Curci:1983an}
G.~Curci, P.~Menotti, and G.~Paffuti, {\it Symanzik's improved lagrangian for
  lattice gauge theory},  {\em Phys. Lett.} {\bf B130} (1983) 205.

\bibitem{Gattringer:2001jf}
C.~Gattringer, R.~Hoffmann, and S.~Schaefer, {\it {Setting the scale for the
  Luescher-Weisz action}},  {\em Phys. Rev.} {\bf D65} (2002) 094503,
  [\href{http://xxx.lanl.gov/abs/hep-lat/0112024}{{\tt hep-lat/0112024}}].

\end{thebibliography}\endgroup
\end{document}